\documentclass[pre,aps,twocolumn,floatfix]{revtex4-2}
\usepackage{graphicx,amssymb,amsmath,ifthen,rotating}
\usepackage{float}
\usepackage{color}
\usepackage{algorithm}
\usepackage{algorithmic}

\begin{document}

\title{An exact density-functional closure for the two-dimensional discrete
wormlike chain}

\author{Benaoumeur Bakhti}
\email{bbakhti@uni-osnabrueck.de,benaoumeur.bakhti@univ-mascara.dz}
\affiliation{Department of Physics, University of Mascara, Mascara 29000, Algeria}

\date{\today}

\begin{abstract}
We formulate an exact density-functional description of the angularly discretized two-dimensional (2D) discrete wormlike chain (DWLC) under tension. The central result is a bond-local closure connecting the nearest-neighbor pair distribution \(C_{i,i+1}^{rs}\) to the single-site angular densities \(\rho_i^r\): $C_{i,i+1}^{rs}C_{i,i+1}^{00}/(C_{i,i+1}^{r0}C_{i,i+1}^{0s})=e^{\bar\beta\,rs}$,
with the bond coupling $\bar\beta=\beta\kappa\varepsilon^2/a$.
This relation follows directly from the connected Boltzmann weight of the quadratic bending interaction and permits exact integration of the entropy functional. Variational minimization then yields coupled self-consistent equations for the one- and two-site angular distributions, which we solve by fixed-point iteration. We establish the equivalence of this density-functional formulation to the exact transfer-matrix solution: the two approaches reproduce the angular marginals and force-extension curves to machine precision. The formulation also recovers the continuum 2D wormlike-chain behavior, including the rigid-rod and random-coil limits of the mean-square end-to-end distance. Using the segment length and persistence length taken directly from Mazur’s short-DNA molecular-dynamics study, without additional fitting, the predicted bend-angle statistics agree with the simulation data within the estimated uncertainty. The closure structure extends naturally to nonharmonic local bending interactions and can also be interpreted inversely, allowing measured nearest-neighbor angular correlations to constrain effective coarse-grained bending potentials. These properties establish a direct connection between conformational statistics and local interactions and provide a density-level framework for treating interacting semiflexible-polymer systems.
\end{abstract}


\maketitle
\section{Introduction}
Semiflexible polymers represent a crucial class of biological and synthetic
macromolecules whose mechanical properties are essential to cellular function
and material design
\cite{Connolly2019,Xu2018,Dai2016,Chen2010,Janke2016,Bartha2000,Prasad2005}.
Double-stranded DNA, F-actin filaments, and microtubules exemplify natural
semiflexible polymers that undergo complex mechanical transformations in
response to cellular processes: DNA bending is essential for gene regulation and
protein binding, actin polymerization controls cell shape and motility, and
microtubule mechanics regulate intracellular transport \cite{Broedersz2014,Dai2016}.
Beyond biology, synthetic semiflexible polymers find applications in
liquid-crystalline materials \cite{Binder2020,Zhang2015} and in optoelectronic
materials \cite{Botiz2010}, where chain stiffness governs device performance.

Experimental methods for probing semiflexible polymer mechanics have advanced
dramatically. Optical and magnetic tweezers apply piconewton forces to
individual chains \cite{Smith1992,Smith1996}, atomic force microscopy images
nanoscale configurations \cite{Wiggins2006}, and molecular dynamics (MD)
simulations now routinely exceed nanosecond timescales \cite{Mazur2007,Midya2019}.
These experiments demand theoretical frameworks that predict force-extension
relations, bend-angle distributions, and configurational entropies.

The classical freely jointed chain model \cite{Kuhn1934} treats polymers as
sequences of independent rigid segments and fails for semiflexible chains where
bending costs substantial energy. The wormlike chain (WLC) model
\cite{Kratky1949,Liao2020,Milstein2013,Cinacchi2008,Heussinger2007,Samuel2002}
represents the polymer as a space curve with quadratic bending energy and
accurately describes DNA, actin, and many other polymers
\cite{Marantan2018,Mazur2007}. Being continuous, the WLC requires path integrals
that rarely admit simple closed forms; the discrete wormlike chain (DWLC) model
\cite{Fiasconaro2023,Marantan2018,Ghosh2009,Wiggins2005,Lamura2001,Nikoubashman2021}
discretizes the chain into rigid segments with angle variables, reducing the
problem to discrete probability distributions.

The present work treats the
two-dimensional inextensible DWLC of a single chain under tension. A single
chain with nearest-neighbor bending coupling is a one-dimensional
nearest-neighbor statistical system, and such systems are solved exactly and in
full generality by the transfer-matrix method: the partition function, free
energy, and all correlation functions follow from a single matrix. Exact
transfer-operator solutions of semiflexible chains in two dimensions are
well established --- Prasad, Hori, and Kondev \cite{Prasad2005} obtain the 2D
force-extension and nematic-field response in terms of Mathieu functions, and
Fiasconaro and Falo \cite{Fiasconaro2023} diagonalize the discrete (extensible)
WLC transfer matrix and cross-check it against Langevin simulation. Marantan and
Mahadevan \cite{Marantan2018} solve the discrete quadratic-energy chain exactly
by treating the interior angles as a multivariate Gaussian. 

W develop here a density-functional closure for the DWLC. In
place of diagonalizing an operator, we express the pair angular distribution
$C_{i,i+1}^{rs}$ (defined precisely in Sec.~\ref{sec:closure}) exactly in terms
of the single-site angular density $\rho_i^r$ through a closure relation, build
an entropy functional by integrating that relation, and obtain the equilibrium
state from a variational principle. For the single chain the resulting
self-consistent equations are equivalent to the transfer-matrix recursion, and
we verify this equivalence numerically to machine precision
(Sec.~\ref{sec:validation}). The motivation for casting the problem as a density
functional is not efficiency for one chain (the transfer matrix is already
elementary there) but extensibility: a closure written in terms of one-body
and two-body densities is the natural object for the interacting many-chain
regime, where the configuration space grows exponentially and the transfer
matrix becomes intractable. In such a many-chain extension the one-body angular
density would describe the local orientational field, while the two-body density
would provide the level at which excluded-volume or other inter-chain
interactions naturally enter. We touch on this direction with a qualitative
mean-field discussion of the isotropic-nematic transition, presented explicitly
as an outlook rather than a controlled result.

Two further features of the closure are worth stating, as they broaden its
scope beyond the specific model solved here. First, the construction is not
intrinsically restricted to the harmonic bending potential considered in this
manuscript: we have also verified its applicability to an anharmonic discrete
wormlike-chain model, demonstrating that the same closure strategy can
accommodate more general local bending interactions; that extension is being
developed separately and will be reported in a forthcoming study \cite{Bakhti2026}. Second, the
closure relates the local two-point angular distribution directly to the
underlying local interaction, and so may be read in either direction --- a point
we return to below Eq.~(\ref{eq:closure}).

The paper is organized as follows.
Section~\ref{sec:model} reformulates the WLC model.
Section~\ref{sec:closure} derives the exact closure relation.
Section~\ref{sec:functional} presents the free-energy functional and the
self-consistent equations. Section~\ref{sec:numerics} describes the
numerical algorithm and how the mean extension and its higher moments are
computed. Section~\ref{sec:validation} establishes the equivalence with, and
validation against, the exact transfer matrix.
Sections~\ref{sec:forceext} and \ref{sec:limits} treat force-extension behavior
and the rigid/flexible limits. Section~\ref{sec:md} compares with the short-DNA
MD data of Mazur, and Sec.~\ref{sec:conclusion} concludes. Lengthy derivations are
collected in the appendices.
\section{Wormlike chain model in two dimensions}\label{sec:model}
\subsection{Continuum formulation}
Consider a single semiflexible polymer in two dimensions, of total contour
length $L$, described as a plane curve $\mathbf{r}(s)$ parametrized by arc length
$s\in[0,L]$. The unit tangent $\mathbf{t}(s)=d\mathbf{r}/ds$ makes an angle
$\theta(s)$ with a fixed axis. The bending energy is
\begin{equation}\label{eq:H_bending}
\mathcal{H}_0 = \frac{\kappa}{2}\int_0^{L}\!ds\left(\frac{d\theta}{ds}\right)^2,
\end{equation}
where $\kappa$ is the bending stiffness (units energy$\times$length). In two
dimensions the tangent-tangent correlation decays as
$\langle\cos[\theta(s)-\theta(0)]\rangle=e^{-s/l_p}$ with persistence length
\begin{equation}\label{eq:l_p}
\,l_p = 2\beta\kappa\,,\qquad \beta=1/(k_BT).
\end{equation}
The factor of two is specific to $d=2$: it arises because a plane curve has a
single bending mode, whereas the more familiar $l_p=\beta\kappa$ holds in three
dimensions, where there are two independent transverse modes. 

For a chain held under tension $F$ --- that is, subjected to equal and
opposite forces $\pm F\hat{\mathbf x}$ at its two ends, so that only the relative
(end-to-end) coordinate enters --- the external energy is
\begin{equation}\label{eq:H_external_cont}
\mathcal{H}_1 = -F\int_0^{L}\!ds\,\cos\theta(s) = -F R_x .
\end{equation}
We emphasize the wording "tension" rather than "a force pulling one end":
a single unbalanced force cannot act on a free chain, and the clamped end (which
does no work) or an applied reaction $-F$ supplies the balance; either picture
gives the same $-F R_x$.

\subsection{Discrete formulation}
\begin{figure}[h]
  \begin{center}
    \includegraphics[width=0.95\columnwidth]{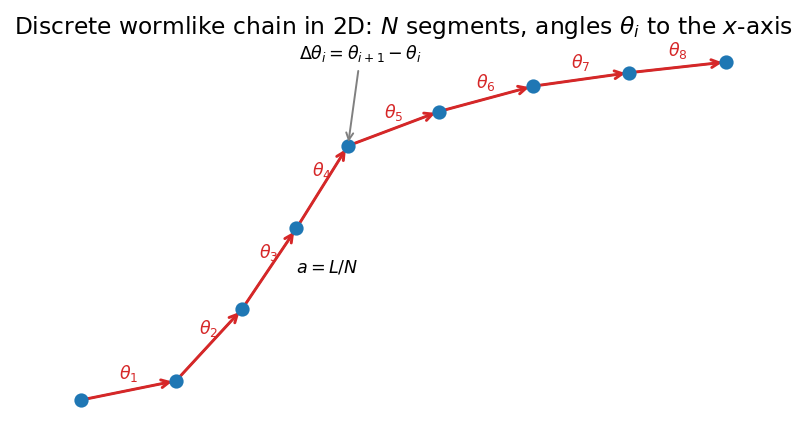}
  \end{center}
\caption{Discrete wormlike chain in two dimensions. A semiflexible polymer of
length $L$ is divided into $N_s$ segments of length $a=L/N_s$; segment $i$ makes
angle $\theta_i$ with the $x$-axis, and the bending energy depends on the
differences $\Delta\theta_i=\theta_{i+1}-\theta_i$. As $a\to0$ the model
converges to the continuum 2D WLC.}
\label{fig:fig1}
\end{figure}

We discretize the chain into $N_s$ segments of length $a=L/N_s$, with
orientation angles $\theta_i$, $i=1,\dots,N_s$ (Fig.~\ref{fig:fig1}).
The bending energy becomes
\begin{equation}\label{eq:H_discrete_bend}
\mathcal{H}_0 = \frac{\kappa}{2a}\sum_{i=1}^{N_s-1}(\theta_{i+1}-\theta_i)^{2},
\end{equation}
the factor $\kappa/(2a)$ ensuring that as $a\to0$ at fixed $\kappa$ the sum
approaches Eq.~(\ref{eq:H_bending}). Under tension,
\begin{equation}\label{eq:H_discrete_ext}
\mathcal{H}_1 = -Fa\sum_{i=1}^{N_s}\cos\theta_i .
\end{equation}

The primary observable is the mean extension along the force,
\begin{equation}\label{eq:R_def}
\langle R\rangle = a\sum_{i=1}^{N_s}\langle\cos\theta_i\rangle .
\end{equation}
At zero force this vanishes identically by rotational symmetry
($\langle\cos\theta_i\rangle=0$), so the relevant zero-field observable is the
mean-square end-to-end distance,
\begin{equation}\label{eq:R2_def}
\langle R^2\rangle = a^2\sum_{i,j}\langle\cos(\theta_i-\theta_j)\rangle,
\end{equation}
which we use for the rigid-rod and random-coil limits in
Sec.~\ref{sec:limits}. The same formalism gives the second moment of the
extension and other observables that can be expressed through one- and two-site
angular correlations; higher moments such as $\langle R^4\rangle$ involve
multi-site correlations beyond the pair level and are not directly accessible at
this order.
\section{Exact closure relation}\label{sec:closure}
\subsection{Angular densities}
We discretize the orientation into $n$ bins, $\theta\to r\varepsilon$ with
$\varepsilon=2\pi/n$ and $r=0,\dots,n-1$. The single-site angular density and the
nearest-neighbor pair distribution are
\begin{align}
\rho_i^r &= P(\theta_i=r\varepsilon),\\
C_{i,i+1}^{rs} &= P(\theta_i=r\varepsilon,\ \theta_{i+1}=s\varepsilon).
\end{align}
(The bin width $\varepsilon$ can be refined toward the continuum by increasing
$n$; the discrete grid is adequate for the applications below.) The two are not
independent: mean-field factorization $C_{i,i+1}^{rs}=\rho_i^r\rho_{i+1}^s$
ignores the bending coupling between adjacent bonds and is uncontrolled. We now
derive the exact relation that replaces it.

\subsection{Closure from the connected bond weight}
The bending energy of the bond $(i,i+1)$ is
\begin{equation}\label{eq:phi_bond}
\phi_{i,i+1}^{rs}=\frac{\kappa}{2a}\,(r-s)^2\varepsilon^2 .
\end{equation}
Alongside $C_{i,i+1}^{rs}$ we introduce three auxiliary distributions in which
one or both of the two angles are pinned to the reference orientation
$\theta=0$:
\begin{align}
C_{i,i+1}^{r0}&=P(\theta_i=r\varepsilon,\theta_{i+1}=0),\\
C_{i,i+1}^{0s}&=P(\theta_i=0,\theta_{i+1}=s\varepsilon),\\
C_{i,i+1}^{00}&=P(\theta_i=0,\theta_{i+1}=0).
\end{align}
Together with $C_{i,i+1}^{rs}$ these are the four two-site objects entering the
closure. Each is a marginal of the full joint distribution and therefore carries
the same single-site ("field") weights coming from the rest of the
chain; those weights cancel in the combination
$C^{rs}C^{00}/(C^{r0}C^{0s})$, leaving only the bond Boltzmann factor. Writing
the marginal of a single bond as $C_{i,i+1}^{rs}\propto
\mathcal{L}_i(r)\,e^{-\beta\phi_{i,i+1}^{rs}}\,\mathcal{R}_{i+1}(s)$, with
$\mathcal{L},\mathcal{R}$ the contributions of the chain to the left and right
(the transfer-matrix messages of Sec.~\ref{sec:validation}), one finds that
$\mathcal{L}_i(r)$ and $\mathcal{R}_{i+1}(s)$ cancel identically and
\begin{equation}\label{eq:closure}
\;\frac{C_{i,i+1}^{rs}\,C_{i,i+1}^{00}}
{C_{i,i+1}^{r0}\,C_{i,i+1}^{0s}}
= \exp\!\big(\bar\beta\,rs\big)\;,
\end{equation}
where the dimensionless bond coupling is
\begin{equation}\label{eq:betabar}
\bar\beta \;\equiv\; \frac{\beta\kappa\,\varepsilon^2}{a}
\;=\; \frac{l_p\,\varepsilon^2}{2a}.
\end{equation}
The exponent is precisely the connected part of the bond energy,
\begin{equation}
-\beta\big(\phi^{rs}-\phi^{r0}-\phi^{0s}+\phi^{00}\big)
=\frac{\beta\kappa\varepsilon^2}{a}\,rs=\bar\beta\,rs,
\end{equation}
obtained from Eq.~(\ref{eq:phi_bond}) using
$(r-s)^2-r^2-s^2=-2rs$. Equation~(\ref{eq:closure}) is exact: it holds for every
bond, independent of the boundary conditions and of the external field, because
the message factors cancel. All factors of the segment length $a$ are retained
explicitly in Eq.~(\ref{eq:betabar}); a full derivation, including the
four-configuration energy bookkeeping, is given in Appendix~\ref{app:closure}.
The continuum limit $a\to0$ at fixed $\kappa$ (hence fixed $l_p$) is taken with
$N_s=L/a\to\infty$; the discrete correlations then approach the continuum WLC
result, as verified in Sec.~\ref{sec:limits}.

Equation~(\ref{eq:closure}) also admits an inverse reading. Within the
assumptions of the present construction, it ties the local nearest-neighbor
two-point angular distribution directly to the underlying local bending
interaction, through the single coupling $\bar\beta$. Thus, while a specified
interaction predicts the corresponding angular correlations (the forward
direction used throughout this paper), measured nearest-neighbor angular
correlations could in principle be used in the reverse direction to infer an
effective local interaction of a semiflexible polymer. We regard this
inverse aspect (available because the theory is expressed directly in terms of
one- and two-point angular densities) as a potentially useful consequence of
the formulation, subject to the caveat that the relation carries the assumptions
of the closure and a residual freedom associated with the one-body factors, and
is not claimed here as a general inversion theorem.

\subsection{Pair correlations in terms of densities}
The auxiliary distributions are fixed by the single-site densities through the
normalization sum rules
\begin{align}
C_{i,i+1}^{r0} &= \rho_i^r - \sum_{l=1}^{n} C_{i,i+1}^{rl},\label{eq:sr1}\\
C_{i,i+1}^{0s} &= \rho_{i+1}^s - \sum_{l=1}^{n} C_{i,i+1}^{ls},\label{eq:sr2}\\
C_{i,i+1}^{00} &= 1 - \sum_{l=1}^{n}(\rho_i^l + \rho_{i+1}^l)
+ \sum_{l,m=1}^{n} C_{i,i+1}^{lm}.\label{eq:sr3}
\end{align}
Substituting into Eq.~(\ref{eq:closure}) gives a closed nonlinear relation for
$C_{i,i+1}^{rs}$ in terms of $\{\rho_i^r\}$ alone, with no approximation.
\section{Free-energy functional and equilibrium}\label{sec:functional}
The constrained free-energy functional of the discretized chain is
\begin{equation}\label{eq:free_energy_full}
\Omega[\rho,C]=\sum_i\sum_{r,s}\phi_{i,i+1}^{rs}C_{i,i+1}^{rs}
- T\,S[\rho,C] + \sum_i\sum_r (V_i^r-\mu)\rho_i^r,
\end{equation}
where $V_i^r$ is an external potential (defined at the discrete angles; for
continuous angles it is interpolated). We work at fixed chain length; $\mu$ is
introduced solely as a Lagrange multiplier enforcing the normalization
$\sum_r\rho_i^r=1$ at each site, so that no polydispersity or grand-canonical
chain exchange is implied. The entropy functional, obtained by integrating the
closure~(\ref{eq:closure}) (Appendix~\ref{app:entropy}), is
\begin{widetext}
\begin{align}\label{eq:entropy}
T S[\rho,C] = &-\sum_{i=1}^{N_s-1}\Bigg[
\sum_{r}\Big(\rho_i^r-\sum_l C_{i,i+1}^{rl}\Big)\ln\Big(\rho_i^r-\sum_l C_{i,i+1}^{rl}\Big)
+\sum_{s}\Big(\rho_{i+1}^s-\sum_l C_{i,i+1}^{ls}\Big)\ln\Big(\rho_{i+1}^s-\sum_l C_{i,i+1}^{ls}\Big)\nonumber\\
&+\Big(1-\sum_l(\rho_i^l+\rho_{i+1}^l)+\sum_{l,m}C_{i,i+1}^{lm}\Big)
\ln\Big(1-\sum_l(\rho_i^l+\rho_{i+1}^l)+\sum_{l,m}C_{i,i+1}^{lm}\Big)\nonumber\\
&+\sum_{r,s}C_{i,i+1}^{rs}\ln C_{i,i+1}^{rs}
-\bar\beta\sum_{r,s} rs\,\big(C_{i,i+1}^{rs}-\rho_i^r\rho_{i+1}^s\big)\Bigg].
\end{align}
\end{widetext}
The last term carries the $rs$ coupling of Eq.~(\ref{eq:closure}); it replaces
the mean-field internal energy exactly. Minimizing $\Omega$,
\begin{equation}\label{eq:variational}
\frac{\delta\Omega}{\delta\rho_i^r}=0,\qquad
\frac{\delta\Omega}{\delta C_{i,i+1}^{rs}}=0,
\end{equation}
yields the self-consistent equations. Variation with respect to
$C_{i,i+1}^{rs}$ reproduces the closure in the equivalent logarithmic form
\begin{align}\label{eq:eqC}
\ln C_{i,i+1}^{rs}=\;&\bar\beta\,rs
+\ln\!\Big(\rho_i^r-\!\sum_l C_{i,i+1}^{rl}\Big)\nonumber\\
&+\ln\!\Big(\rho_{i+1}^s-\!\sum_l C_{i,i+1}^{ls}\Big)\nonumber\\
&-\ln\!\Big(1-\!\sum_l(\rho_i^l+\rho_{i+1}^l)+\!\sum_{l,m}C_{i,i+1}^{lm}\Big),
\end{align}
and variation with respect to $\rho_i^r$ gives the density equation
\begin{equation}\label{eq:eqrho}
\beta(\mu-V_i^r)=\ln\rho_i^r-\ln\Big(1-\sum_l\rho_i^l\Big)
+\Lambda_i^r[\rho,C],
\end{equation}
where $\Lambda_i^r$ collects the correlation contributions from the two bonds
adjacent to site $i$ (explicit form in Appendix~\ref{app:entropy}).
Equations~(\ref{eq:eqC})--(\ref{eq:eqrho}) are the exact stationarity conditions
of the functional; they are transparent only once the numerical scheme that
solves them is in place, which we describe next.
\section{Numerical scheme}\label{sec:numerics}
The self-consistent equations are solved by fixed-point iteration. Because the
scheme is what turns the exact relations into observables, we give it in the main
text.
\begin{algorithm}[H]
\caption{Self-consistent solution of the 2D DWLC}
\begin{algorithmic}[1]
\STATE Initialize $\rho_i^r=1/n$; set tolerance $\epsilon=10^{-12}$.
\STATE \textbf{Forward sweep:} propagate messages
$\mathcal{L}_{i}(r)=\big[\sum_{s}\mathcal{L}_{i-1}(s)\,e^{-\beta\phi^{sr}}\big]\,
e^{\,f\cos(r\varepsilon)}$, $f=\beta F a$, normalizing at each step.
\STATE \textbf{Backward sweep:} propagate $\mathcal{R}_i(r)$ analogously from the
far end.
\STATE \textbf{Densities:} $\rho_i^r\propto \mathcal{L}_i(r)\,\mathcal{R}_i(r)/e^{\,f\cos(r\varepsilon)}$,
normalized to $\sum_r\rho_i^r=1$.
\STATE \textbf{Pair correlations:}
$C_{i,i+1}^{rs}\propto \mathcal{L}_i(r)\,e^{-\beta\phi^{rs}}\,\mathcal{R}_{i+1}(s)$,
normalized to $\sum_{rs}C^{rs}=1$; these satisfy Eqs.~(\ref{eq:closure}) and
(\ref{eq:sr1})--(\ref{eq:sr3}) identically.
\STATE \textbf{Convergence:} stop when $\|\rho^{\text{new}}-\rho^{\text{old}}\|_\infty<\epsilon$.
\STATE \textbf{Observables:} evaluate
$\langle R\rangle=a\sum_i\sum_r\cos(r\varepsilon)\rho_i^r$; the mean-square size
$\langle R^2\rangle$ from Eq.~(\ref{eq:R2_def}) using the pair marginals; and the
order parameter $S=\sum_r\cos(2r\varepsilon)\rho^r$.
\end{algorithmic}
\end{algorithm}
For a homogeneous field the forward-backward sweep converges in a single pass;
with a site-dependent field a handful of iterations suffice. The cost is
$O(N_s n^2)$ per iteration, making the calculation inexpensive for the
discretizations used here. The second moment follows from the same one- and
two-site densities: the variance $\langle R^2\rangle-\langle R\rangle^2$ is
assembled from the pair marginals $\langle\cos(\theta_i-\theta_j)\rangle$, which
the transfer-matrix messages deliver at no extra asymptotic cost.
\section{Equivalence with, and validation against, the transfer matrix}
\label{sec:validation}
For the single 2D chain, the partition function
\begin{equation}
\mathcal{Z}=\!\!\sum_{r_1\ldots r_{N_s}}\!\!\prod_i v_{r_i}\!\prod_i B_{r_ir_{i+1}},
\ \ v_r=e^{f\cos(r\varepsilon)},\ B_{rs}=e^{-\beta\phi^{rs}}
\end{equation}
is evaluated exactly by the symmetric transfer matrix
$T=V^{1/2}BV^{1/2}$ with $V=\mathrm{diag}(v_r)$, giving
$\mathcal{Z}=u^{\mathsf T}T^{N_s-1}u$ ($u_r=\sqrt{v_r}$) and the exact single-site
marginals $\rho_i^r\propto(u^{\mathsf T}T^{i-1})_r(T^{N_s-i}u)_r$. This is the
elementary exact solution referred to in the Introduction. The forward-backward
messages of Sec.~\ref{sec:numerics} are the partial products
$u^{\mathsf T}T^{i-1}$ and $T^{N_s-i}u$; consequently the self-consistent
density-functional solution and the transfer matrix are two views of the same
computation for the single chain.

\begin{figure*}[htb]
  \begin{center}
    \includegraphics[width=0.9\textwidth]{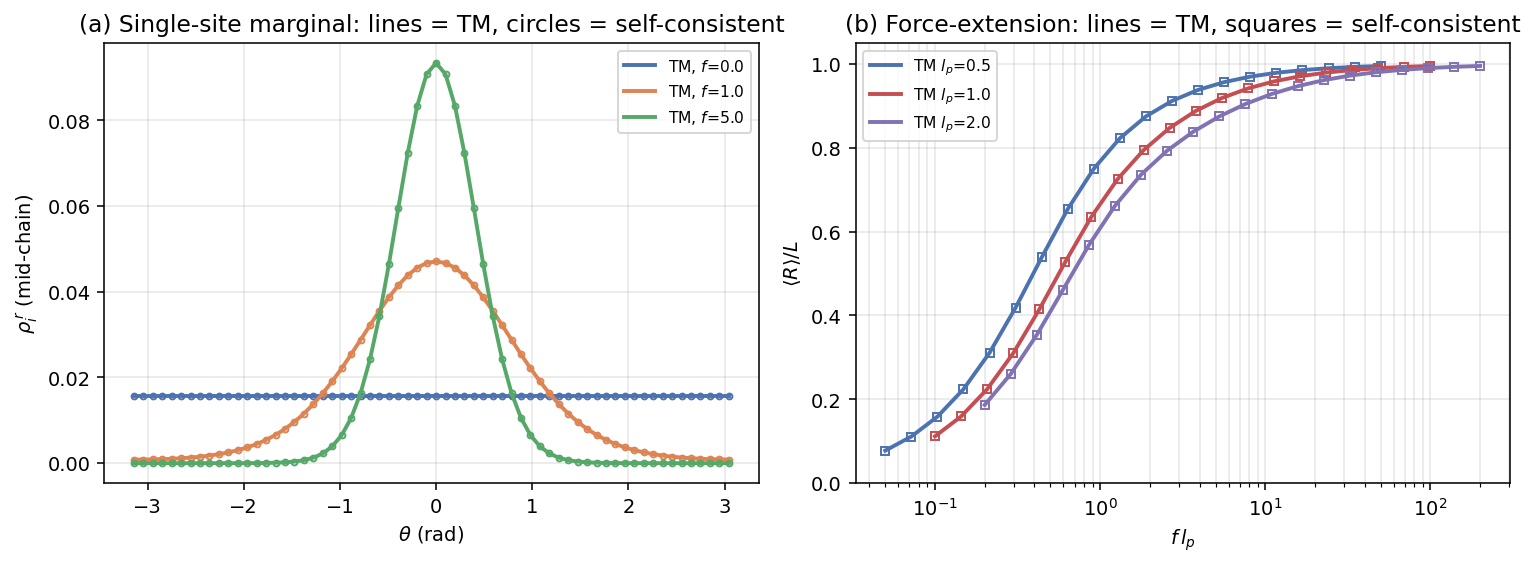}
  \end{center}
\caption{Validation of the self-consistent density functional against the exact
transfer matrix. \textit{(a)} Mid-chain angular marginal $\rho_i^r$ for
$l_p=1$, $N_s=20$ at three tensions $f=\beta Fa$; lines are the transfer matrix,
open circles the self-consistent solution. At $f=0$ the marginal is flat
(rotational symmetry). \textit{(b)} Force-extension $\langle R\rangle/L$ for
$l_p=0.5,1,2$; lines are the transfer matrix, squares the self-consistent
solution. The two methods coincide to machine precision.}
\label{fig:fig6}
\end{figure*}

We confirm this numerically. Solving the same model in its transfer-matrix
formulation and comparing with the density-functional self-consistent solution,
we find that the single-site marginals agree to $8\times10^{-17}$ (i.e., to
machine precision) across all persistence lengths and forces tested, and the
extensions coincide to all quoted digits (Table~\ref{tab:tm}).
Figure~\ref{fig:fig6} overlays the two solutions for the mid-chain angular
marginal and for the force-extension curve; the circles (self-consistent) sit
exactly on the transfer-matrix lines. Since, as noted above, the two are the same
forward-backward recursion written in different ways, this is a consistency check
that the self-consistent implementation faithfully realizes the exact
single-chain result, rather than an independent route to it.
\begin{table}[H]
\begin{center}
\begin{tabular}{|c|c|c|c|c|}
\hline
$l_p/a$ & $f=\beta Fa$ & $\langle R\rangle/N_sa$ (SC) & $\langle R\rangle/N_sa$ (TM) & $\max_i|\Delta\rho_i|$\\
\hline
0.5 & 0 & $0.0000$ & $0.0000$ & $1{\times}10^{-17}$\\
0.5 & 1 & $0.5760$ & $0.5760$ & $3{\times}10^{-17}$\\
1.0 & 0 & $0.0000$ & $0.0000$ & $1{\times}10^{-17}$\\
1.0 & 1 & $0.6638$ & $0.6638$ & $4{\times}10^{-17}$\\
1.0 & 5 & $0.9116$ & $0.9116$ & $7{\times}10^{-17}$\\
2.0 & 1 & $0.7522$ & $0.7522$ & $4{\times}10^{-17}$\\
\hline
\end{tabular}
\caption{Self-consistent density-functional solution (SC) versus exact transfer
matrix (TM) for $N_s=20$, $n=64$. The mean extension agrees to all quoted
digits and the single-site marginals agree to machine precision. Note that the
zero-field extension vanishes by rotational symmetry.}
\label{tab:tm}
\end{center}
\end{table}
\section{Force-extension under tension}\label{sec:forceext}
\begin{figure*}[t]
\centering
\includegraphics[width=0.95\textwidth]{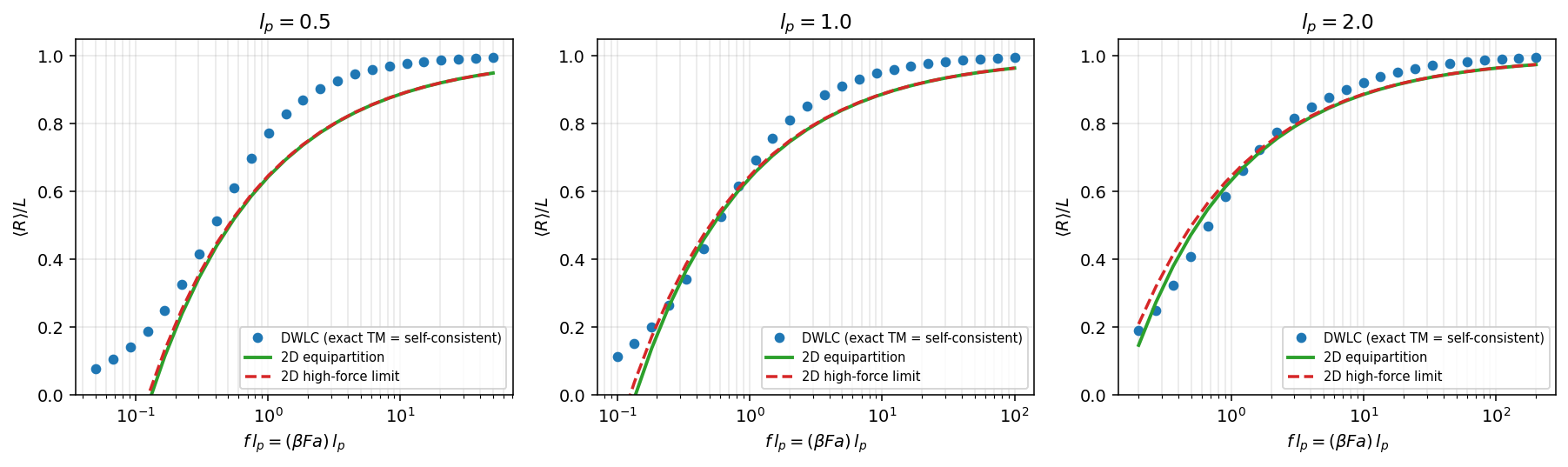}
\caption{Force-extension of the 2D DWLC for $l_p=0.5,1,2$ ($N_s=40$).
Circles: exact discrete result (self-consistent $\equiv$ transfer matrix).
Solid: 2D equipartition formula, Eq.~(\ref{eq:eq2D}). Dashed: 2D high-force
limit, Eq.~(\ref{eq:hf2D}). Agreement is at the few-percent level for
$fl_p\gtrsim1$; deviations at small $fl_p$ reflect finite discretization.}
\label{fig:fig3}
\end{figure*}

Under tension the chain aligns with the force and $\langle R\rangle$ grows toward
$L$. For comparison, we consider the continuum two-dimensional wormlike-chain (WLC) model. A Gaussian
(equipartition) treatment of the transverse fluctuations in two dimensions gives,
for a chain of length $L$,
\begin{equation}\label{eq:eq2D}
\frac{\langle R\rangle}{L}=1-\frac{1}{4\sqrt{\beta\kappa\,f}}
\coth\!\Big(L\sqrt{f/\beta\kappa}\Big)-\frac{1}{4fL},
\end{equation}
with $\beta\kappa=l_p/2$, whose high-force limit is
\begin{equation}\label{eq:hf2D}
\frac{\langle R\rangle}{L}\simeq 1-\frac{1}{4\sqrt{\beta\kappa f}}
=1-\frac{1}{2\sqrt{2\,f l_p}}\qquad(fl_p\gg1).
\end{equation}
Figure~\ref{fig:fig3} compares the exact discrete result (which, per
Sec.~\ref{sec:validation}, is simultaneously the self-consistent and the
transfer-matrix solution) with Eqs.~(\ref{eq:eq2D}) and (\ref{eq:hf2D}). In the
high-force regime $fl_p\gtrsim1$ the exact curve tracks the 2D equipartition
formula to within a few percent; the residual deviation is a genuine
discretization/finite-length effect and decreases as $n$ and $N_s$ grow.
\section{Rigid and flexible limits at zero field}\label{sec:limits}
\begin{figure*}[t]
  \begin{center}
    \includegraphics[width=0.95\textwidth]{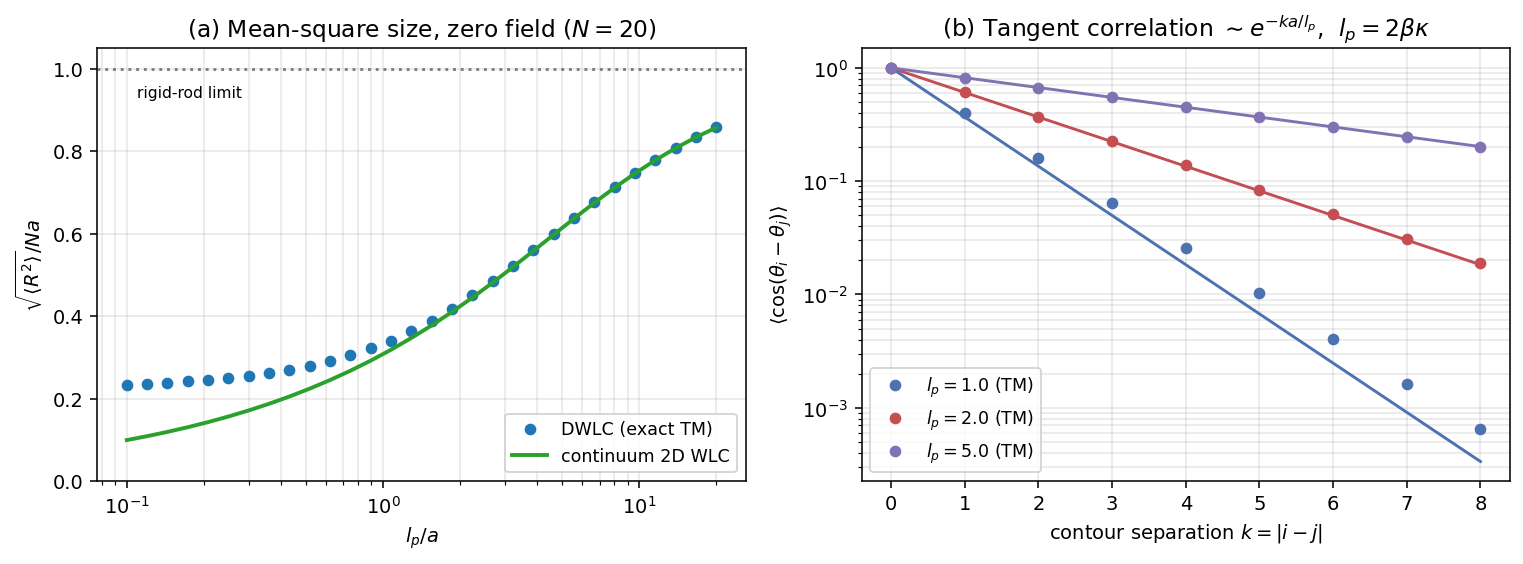}
  \end{center}
\caption{Zero-field behavior of the 2D DWLC. \textit{(a)} Root-mean-square size
$\sqrt{\langle R^2\rangle}/N_sa$ versus persistence length; exact discrete result
(circles) and continuum 2D WLC (line). The curve interpolates between the
random-coil regime and the rigid-rod plateau at unity.
\textit{(b)} Tangent correlation versus contour separation for
$l_p=1,2,5$; points are exact, lines are $e^{-|i-j|a/l_p}$, confirming
$l_p=2\beta\kappa$.}
\label{fig:fig2}
\end{figure*}

Because $\langle R\rangle=0$ at zero field, the limiting behavior is read from
the mean-square size~(\ref{eq:R2_def}) and the tangent correlation.
Figure~\ref{fig:fig2}(a) shows $\sqrt{\langle R^2\rangle}/N_sa$ versus $l_p/a$:
it rises from the random-coil regime, where
$\langle R^2\rangle\to2l_pL$ (Gaussian scaling, $\sqrt{\langle
R^2\rangle}\sim\sqrt{N_s}$), toward the rigid-rod plateau
$\sqrt{\langle R^2\rangle}/N_sa\to1$ as $l_p/a\to\infty$. The exact discrete data
follow the continuum 2D WLC prediction
$\langle R^2\rangle=2l_pL[1-(l_p/L)(1-e^{-L/l_p})]$ over the stiff and
intermediate range, deviating only in the strongly flexible regime where the
large per-bond angles resolve the discreteness of the model.
Figure~\ref{fig:fig2}(b) shows the tangent correlation
$\langle\cos(\theta_i-\theta_j)\rangle$ decaying as $e^{-|i-j|a/l_p}$; the
discrete points coincide with the continuum exponential for $l_p\gtrsim2a$ and
lie slightly above it for $l_p=a$, again a controlled discretization effect. Both
panels confirm the 2D relation $l_p=2\beta\kappa$.
\section{Comparison with molecular-dynamics data}\label{sec:md}
The all-atom MD study of short DNA by Mazur \cite{Mazur2007} provides a
quantitative benchmark for bend-angle statistics. That work measured bend-angle
distributions for AT-alternating fragments of $4$--$24$ base pairs (bp) and found
them nearly linear in WLC-linearized coordinates, $\ln P$ versus $1-\cos\theta$,
with a persistence length $l_p\approx80$~nm.

To compare, we set the parameters directly from the MD system rather than fitting
them: rise per base pair $a=0.34$~nm, persistence length $l_p=80$~nm (hence
$\beta\kappa=l_p/2=40$~nm and $\bar\beta=l_p\varepsilon^2/2a$ for the chosen
$n$), and fragment length $L=N_{\rm bp}\,a$. With these fixed values the WLC bend
distribution is $P(\theta)\propto\exp[-(l_p/2L)(1-\cos\theta)]$, and the mean
bend angle in the Gaussian regime is
$\langle\theta^2\rangle^{1/2}=(2L/l_p)^{1/2}$. Table~\ref{tab:tab1} lists the
resulting mean bend angles alongside the MD values read from Ref.~\cite{Mazur2007},
with an estimated reading/statistical uncertainty of $\pm0.5^\circ$ on the MD
figures. The discrete model reproduces the MD trend across the whole range within
this uncertainty.
\begin{table}[H]
\begin{center}
\begin{tabular}{|c|c|c|c|}
\hline
Fragment & $L$ (nm) & $\langle\theta\rangle$ (model) & $\langle\theta\rangle$ (MD) \\
\hline
4 bp  & 1.36 & $10.6^\circ$ & $10.6^\circ\pm0.5^\circ$ \\
6 bp  & 2.04 & $12.9^\circ$ & $12.9^\circ\pm0.5^\circ$ \\
8 bp  & 2.72 & $14.9^\circ$ & $14.9^\circ\pm0.5^\circ$ \\
10 bp & 3.40 & $16.7^\circ$ & $16.7^\circ\pm0.5^\circ$ \\
12 bp & 4.08 & $18.3^\circ$ & $18.3^\circ\pm0.5^\circ$ \\
\hline
\end{tabular}
\caption{Mean bend angle of short DNA fragments: discrete WLC model (parameters
$a=0.34$~nm, $l_p=80$~nm taken directly from the MD system) versus the MD data. The MD uncertainty is an estimate of the reading and
statistical error. Model and simulation agree within this uncertainty across the
range.}\label{tab:tab1}
\end{center}
\end{table}
\begin{figure*}[t]
  \begin{center}
    \includegraphics[width=0.95\textwidth]{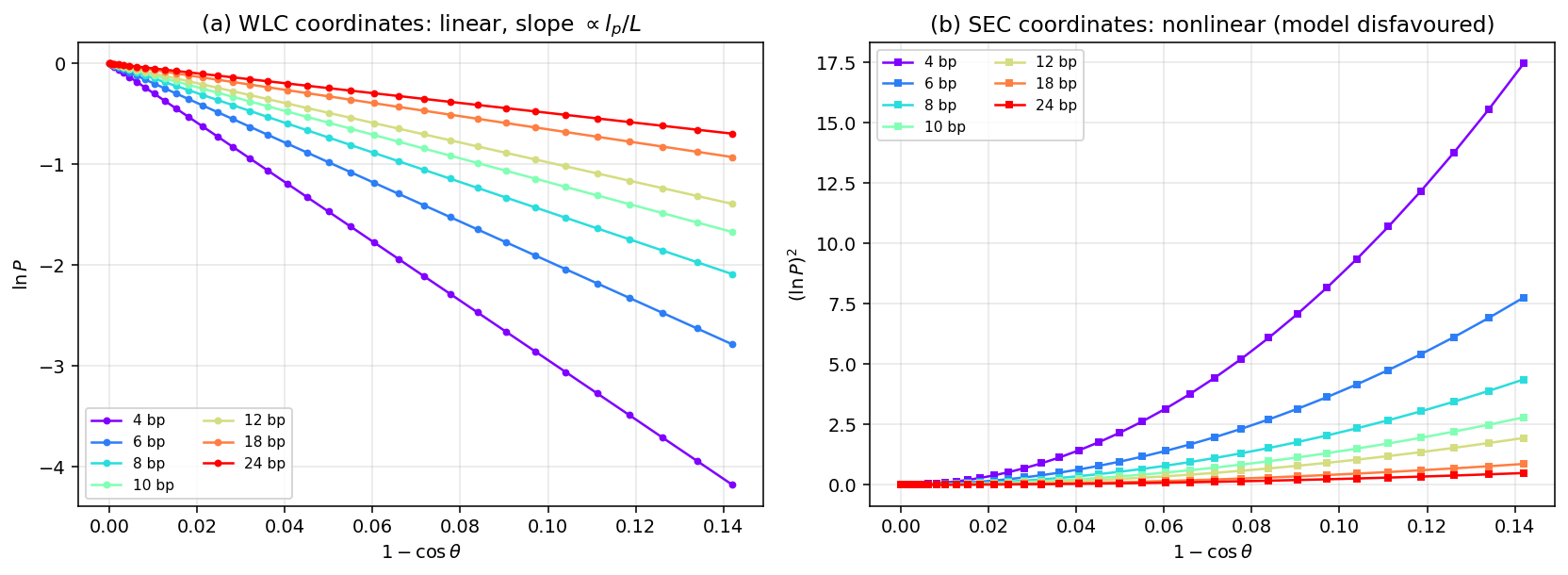}
  \end{center}
\caption{Bend-angle statistics of short DNA. \textit{(a)} WLC-linearized
coordinates $\ln P$ versus $1-\cos\theta$ for fragments of 4--24 bp; the curves
are straight with a common slope, consistent with $l_p\approx80$~nm.
\textit{(b)} The same data in SEC coordinates $(\ln P)^2$ versus $1-\cos\theta$
are curved, disfavoring the SEC model.}
\label{fig:fig5}
\end{figure*}
Figure~\ref{fig:fig5} shows the linearization test. In WLC coordinates (panel (a)) the distributions are straight with a common slope
$\propto l_p/L$, consistent with $l_p\approx80$~nm across all lengths; in the
subelastic-chain (SEC) coordinates $(\ln P)^2$ versus $1-\cos\theta$ (panel (b))
they are visibly curved, disfavoring the SEC form. The small upward deviations at
the largest angles for the shortest fragments (4--10 bp) are consistent with
Mazur's attribution to $\approx8$-bp minor-groove modulations, i.e., structural
heterogeneity superimposed on the WLC baseline, and are not a failure of the
underlying quadratic-bending description.
\section{Conclusion}\label{sec:conclusion}
We have formulated the two-dimensional discrete wormlike chain as a density
functional built on an exact closure relation,
Eq.~(\ref{eq:closure}), between the pair angular distribution and the single-site
angular density. The closure follows transparently from the connected part of the
quadratic bending energy, with the bond coupling
$\bar\beta=\beta\kappa\varepsilon^2/a$ carrying all factors of the segment length
explicitly. It permits an exact integration of the entropy functional and yields
self-consistent equations that we solve by a fast forward-backward iteration.

For a single chain these equations are equivalent to an elementary transfer
matrix, and we have verified the equivalence to machine precision. Using the correct 2D conventions throughout ($l_p=2\beta\kappa$;
2D force-extension asymptotics; nematic order parameter $\langle\cos2\theta\rangle$),
we recover the random-coil and rigid-rod limits through $\langle R^2\rangle$,
reproduce the 2D continuum force-extension law within a few percent in the
high-force regime, and match the short-DNA MD bend-angle data of Mazur within
its uncertainty.

The purpose of the density-functional formulation is extensibility. Because it is
written in terms of one- and two-body densities, it is the natural language for
interacting many-chain problems (nematic ordering, confinement, networks)
where the configuration space is exponentially large and transfer-matrix methods
become impractical. The natural next step is to extend the two-site closure from
nearest-neighbor bonds along a single contour to pairs of segments on different
chains, so that excluded-volume and other inter-chain interactions enter directly
at the two-body level. Such an extension would place the isotropic-nematic
transition of stiff-chain solutions (characterized in two dimensions by the
order parameter $\langle\cos 2\theta\rangle$) within reach of the same
formalism. A simple Onsager-type mean-field estimate indicates that a transition
of this kind should occur, but obtaining it from the closure, rather than
by appending an interaction ansatz by hand, remains open; and because mean-field
orientational order is itself delicate in two dimensions, a controlled treatment
must also confront the role of fluctuations. We regard the construction of a
genuine multi-chain closure as the central task ahead.

The scope of the present treatment should be stated plainly in two respects. First, the entire treatment is
two-dimensional; the physically more important three-dimensional chain, where the
orientation lives on the sphere, the pair object carries $2(d-1)$ indices and is
no longer a symmetric square matrix, and the number of closure relations must be
re-examined, is left for a separate study. Second, we are less sanguine than one
might hope about long-ranged intra-chain interactions: for polyelectrolytes the
effective stiffness arises from spatial Coulomb couplings between all
monomer pairs, which are not nearest-neighbor along the contour and cannot be
absorbed into a bond closure without additional, more complex interaction graphs.
Extending the framework in that direction, if possible at all, will require
genuinely new elements beyond those developed here.

\appendix
\section{Derivation of the closure relation}\label{app:closure}
Consider the bond $(i,i+1)$ with the remainder of the chain integrated out. Its
joint marginal is
$C_{i,i+1}^{rs}=\mathcal{Z}^{-1}\mathcal{L}_i(r)\,e^{-\beta\phi^{rs}_{i,i+1}}\,\mathcal{R}_{i+1}(s)$,
where $\mathcal{L}_i(r)$ ($\mathcal{R}_{i+1}(s)$) is the summed Boltzmann weight
of the sub-chain to the left of $i$ (right of $i+1$) with $\theta_i=r\varepsilon$
($\theta_{i+1}=s\varepsilon$) fixed. The four objects
$C^{rs},C^{r0},C^{0s},C^{00}$ share the identical factors $\mathcal{L}_i$ and
$\mathcal{R}_{i+1}$ evaluated at the same pinned angles, so in the ratio
\begin{equation}
\frac{C_{i,i+1}^{rs}C_{i,i+1}^{00}}{C_{i,i+1}^{r0}C_{i,i+1}^{0s}}
=\frac{e^{-\beta\phi^{rs}}e^{-\beta\phi^{00}}}{e^{-\beta\phi^{r0}}e^{-\beta\phi^{0s}}}
=e^{-\beta(\phi^{rs}+\phi^{00}-\phi^{r0}-\phi^{0s})}
\end{equation}
the message factors cancel exactly. With
$\phi^{rs}=(\kappa/2a)(r-s)^2\varepsilon^2$ and $\phi^{00}=0$,
\begin{equation}
\phi^{rs}+\phi^{00}-\phi^{r0}-\phi^{0s}
=\frac{\kappa\varepsilon^2}{2a}\big[(r-s)^2-r^2-s^2\big]
=-\frac{\kappa\varepsilon^2}{a}\,rs,
\end{equation}
so that the ratio equals $e^{\bar\beta rs}$ with
$\bar\beta=\beta\kappa\varepsilon^2/a=l_p\varepsilon^2/2a$, which is
Eq.~(\ref{eq:closure}). The cancellation of $\mathcal{L}_i,\mathcal{R}_{i+1}$ is
why the closure is exact and independent of the external field and boundary
conditions.

\section{Entropy functional and density equation}\label{app:entropy}
Taking the logarithm of the closure and using the sum
rules~(\ref{eq:sr1})--(\ref{eq:sr3}) gives
\begin{align}
\ln C_{i,i+1}^{rs}=\;&\bar\beta\,rs
+\ln\Big(\rho_i^r-\sum_l C_{i,i+1}^{rl}\Big)\nonumber\\
&+\ln\Big(\rho_{i+1}^s-\sum_l C_{i,i+1}^{ls}\Big)\nonumber\\
&-\ln\Big(1-\sum_l(\rho_i^l+\rho_{i+1}^l)+\sum_{l,m}C_{i,i+1}^{lm}\Big).
\end{align}
Integrating $T S=\sum_{i}\sum_{r,s}\int\phi^{rs}_{i,i+1}\,dC^{rs}_{i,i+1}$ term by
term with $\int\ln(a+x)\,dx=(a+x)[\ln(a+x)-1]$ produces
Eq.~(\ref{eq:entropy}). Variation of $\Omega$ with respect to $\rho_i^r$ gives
Eq.~(\ref{eq:eqrho}) with
\begin{equation}
\Lambda_i^r[\rho,C]=-\!\!\sum_{\langle j\rangle=i\pm1}\!\!
\ln\!\frac{1-\sum_l(\rho_i^l+\rho_j^l)+\sum_{l,m}C^{lm}}{\rho_i^r-\sum_lC^{rl}},
\end{equation}
the sum running over the bonds adjacent to site $i$.

\nocite{REVTEX42Control,apsrev42Control}
\bibliographystyle{apsrev4-2}
\bibliography{mybib}

\begin{thebibliography}{30}%
\makeatletter
\providecommand \@ifxundefined [1]{%
 \@ifx{#1\undefined}
}%
\providecommand \@ifnum [1]{%
 \ifnum #1\expandafter \@firstoftwo
 \else \expandafter \@secondoftwo
 \fi
}%
\providecommand \@ifx [1]{%
 \ifx #1\expandafter \@firstoftwo
 \else \expandafter \@secondoftwo
 \fi
}%
\providecommand \natexlab [1]{#1}%
\providecommand \enquote  [1]{``#1''}%
\providecommand \bibnamefont  [1]{#1}%
\providecommand \bibfnamefont [1]{#1}%
\providecommand \citenamefont [1]{#1}%
\providecommand \href@noop [0]{\@secondoftwo}%
\providecommand \href [0]{\begingroup \@sanitize@url \@href}%
\providecommand \@href[1]{\@@startlink{#1}\@@href}%
\providecommand \@@href[1]{\endgroup#1\@@endlink}%
\providecommand \@sanitize@url [0]{\catcode `\\12\catcode `\$12\catcode
  `\&12\catcode `\#12\catcode `\^12\catcode `\_12\catcode `\%12\relax}%
\providecommand \@@startlink[1]{}%
\providecommand \@@endlink[0]{}%
\providecommand \url  [0]{\begingroup\@sanitize@url \@url }%
\providecommand \@url [1]{\endgroup\@href {#1}{\urlprefix }}%
\providecommand \urlprefix  [0]{URL }%
\providecommand \Eprint [0]{\href }%
\providecommand \doibase [0]{https://doi.org/}%
\providecommand \selectlanguage [0]{\@gobble}%
\providecommand \bibinfo  [0]{\@secondoftwo}%
\providecommand \bibfield  [0]{\@secondoftwo}%
\providecommand \translation [1]{[#1]}%
\providecommand \BibitemOpen [0]{}%
\providecommand \bibitemStop [0]{}%
\providecommand \bibitemNoStop [0]{.\EOS\space}%
\providecommand \EOS [0]{\spacefactor3000\relax}%
\providecommand \BibitemShut  [1]{\csname bibitem#1\endcsname}%
\let\auto@bib@innerbib\@empty
\bibitem [{\citenamefont {Connolly}\ \emph {et~al.}(2019)\citenamefont
  {Connolly}, \citenamefont {Seow}, \citenamefont {Fenati},\ and\ \citenamefont
  {Ellis}}]{Connolly2019}%
  \BibitemOpen
  \bibfield  {author} {\bibinfo {author} {\bibfnamefont {A.~R.}\ \bibnamefont
  {Connolly}}, \bibinfo {author} {\bibfnamefont {N.}~\bibnamefont {Seow}},
  \bibinfo {author} {\bibfnamefont {R.~A.}\ \bibnamefont {Fenati}},\ and\
  \bibinfo {author} {\bibfnamefont {A.~V.}\ \bibnamefont {Ellis}},\ }\bibfield
  {title} {\bibinfo {title} {{DNA} nanostructures},\ }in\ \href@noop {} {\emph
  {\bibinfo {booktitle} {Comprehensive Nanoscience and Nanotechnology}}},\
  \bibinfo {editor} {edited by\ \bibinfo {editor} {\bibfnamefont {D.~L.}\
  \bibnamefont {Andrews}}, \bibinfo {editor} {\bibfnamefont {R.~H.}\
  \bibnamefont {Lipson}},\ and\ \bibinfo {editor} {\bibfnamefont
  {T.}~\bibnamefont {Nann}}}\ (\bibinfo  {publisher} {Academic Press},\
  \bibinfo {address} {Oxford},\ \bibinfo {year} {2019})\ \bibinfo {edition}
  {2nd}\ ed.,\ pp.\ \bibinfo {pages} {1--26}\BibitemShut {NoStop}%
\bibitem [{\citenamefont {Xu}\ and\ \citenamefont {Jiang}(2018)}]{Xu2018}%
  \BibitemOpen
  \bibfield  {author} {\bibinfo {author} {\bibfnamefont {X.}~\bibnamefont
  {Xu}}\ and\ \bibinfo {author} {\bibfnamefont {Y.}~\bibnamefont {Jiang}},\
  }\bibfield  {title} {\bibinfo {title} {Applications of worm-like chain model
  in polymer physics},\ }\href@noop {} {\bibfield  {journal} {\bibinfo
  {journal} {Int. J. Mod. Phys. B}\ }\textbf {\bibinfo {volume} {32}},\
  \bibinfo {pages} {1840006} (\bibinfo {year} {2018})}\BibitemShut {NoStop}%
\bibitem [{\citenamefont {Dai}\ \emph {et~al.}(2016)\citenamefont {Dai},
  \citenamefont {Renner},\ and\ \citenamefont {Doyle}}]{Dai2016}%
  \BibitemOpen
  \bibfield  {author} {\bibinfo {author} {\bibfnamefont {L.}~\bibnamefont
  {Dai}}, \bibinfo {author} {\bibfnamefont {C.~B.}\ \bibnamefont {Renner}},\
  and\ \bibinfo {author} {\bibfnamefont {P.~S.}\ \bibnamefont {Doyle}},\
  }\bibfield  {title} {\bibinfo {title} {The polymer physics of single dna
  confined in nanochannels},\ }\href@noop {} {\bibfield  {journal} {\bibinfo
  {journal} {Adv. Colloid Interface Sci.}\ }\textbf {\bibinfo {volume} {232}},\
  \bibinfo {pages} {80} (\bibinfo {year} {2016})},\ \bibinfo {note}
  {proceedings from the International Workshop on Polyelectrolytes in
  Chemistry, Biology and Technology}\BibitemShut {NoStop}%
\bibitem [{\citenamefont {Chen}\ \emph {et~al.}(2010)\citenamefont {Chen},
  \citenamefont {Wedemeyer},\ and\ \citenamefont {Lapidus}}]{Chen2010}%
  \BibitemOpen
  \bibfield  {author} {\bibinfo {author} {\bibfnamefont {Y.}~\bibnamefont
  {Chen}}, \bibinfo {author} {\bibfnamefont {W.~J.}\ \bibnamefont
  {Wedemeyer}},\ and\ \bibinfo {author} {\bibfnamefont {L.~J.}\ \bibnamefont
  {Lapidus}},\ }\bibfield  {title} {\bibinfo {title} {A general polymer model
  of unfolded proteins under folding conditions},\ }\href@noop {} {\bibfield
  {journal} {\bibinfo  {journal} {J. Phys. Chem. B}\ }\textbf {\bibinfo
  {volume} {114}},\ \bibinfo {pages} {15969} (\bibinfo {year}
  {2010})}\BibitemShut {NoStop}%
\bibitem [{\citenamefont {Janke}\ and\ \citenamefont
  {Marenz}(2016)}]{Janke2016}%
  \BibitemOpen
  \bibfield  {author} {\bibinfo {author} {\bibfnamefont {W.}~\bibnamefont
  {Janke}}\ and\ \bibinfo {author} {\bibfnamefont {M.}~\bibnamefont {Marenz}},\
  }\bibfield  {title} {\bibinfo {title} {Stable knots in the phase diagram of
  semiflexible polymers: A topological order parameter},\ }\href@noop {}
  {\bibfield  {journal} {\bibinfo  {journal} {J. Phys.: Conference Series}\
  }\textbf {\bibinfo {volume} {750}},\ \bibinfo {pages} {012006} (\bibinfo
  {year} {2016})}\BibitemShut {NoStop}%
\bibitem [{\citenamefont {Bartha}\ \emph {et~al.}(2000)\citenamefont {Bartha},
  \citenamefont {Bog\'ar}, \citenamefont {Peeters}, \citenamefont
  {Van~Alsenoy},\ and\ \citenamefont {Van~Doren}}]{Bartha2000}%
  \BibitemOpen
  \bibfield  {author} {\bibinfo {author} {\bibfnamefont {F.}~\bibnamefont
  {Bartha}}, \bibinfo {author} {\bibfnamefont {F.}~\bibnamefont {Bog\'ar}},
  \bibinfo {author} {\bibfnamefont {A.}~\bibnamefont {Peeters}}, \bibinfo
  {author} {\bibfnamefont {C.}~\bibnamefont {Van~Alsenoy}},\ and\ \bibinfo
  {author} {\bibfnamefont {V.}~\bibnamefont {Van~Doren}},\ }\bibfield  {title}
  {\bibinfo {title} {Density-functional calculations of the elastic properties
  of some polymer chains},\ }\href@noop {} {\bibfield  {journal} {\bibinfo
  {journal} {Phys. Rev. B}\ }\textbf {\bibinfo {volume} {62}},\ \bibinfo
  {pages} {10142} (\bibinfo {year} {2000})}\BibitemShut {NoStop}%
\bibitem [{\citenamefont {Prasad}\ \emph {et~al.}(2005)\citenamefont {Prasad},
  \citenamefont {Hori.},\ and\ \citenamefont {Kondev}}]{Prasad2005}%
  \BibitemOpen
  \bibfield  {author} {\bibinfo {author} {\bibfnamefont {A.}~\bibnamefont
  {Prasad}}, \bibinfo {author} {\bibfnamefont {Y.}~\bibnamefont {Hori.}},\ and\
  \bibinfo {author} {\bibfnamefont {J.}~\bibnamefont {Kondev}},\ }\bibfield
  {title} {\bibinfo {title} {Elasticity of semiflexible polymers in two
  dimensions},\ }\href@noop {} {\bibfield  {journal} {\bibinfo  {journal}
  {Phys. Rev. E}\ }\textbf {\bibinfo {volume} {72}},\ \bibinfo {pages} {041918}
  (\bibinfo {year} {2005})}\BibitemShut {NoStop}%
\bibitem [{\citenamefont {Broedersz}\ and\ \citenamefont
  {MacKintosh}(2014)}]{Broedersz2014}%
  \BibitemOpen
  \bibfield  {author} {\bibinfo {author} {\bibfnamefont {C.~P.}\ \bibnamefont
  {Broedersz}}\ and\ \bibinfo {author} {\bibfnamefont {F.~C.}\ \bibnamefont
  {MacKintosh}},\ }\bibfield  {title} {\bibinfo {title} {Modeling semiflexible
  polymer networks},\ }\href@noop {} {\bibfield  {journal} {\bibinfo  {journal}
  {Rev. Mod. Phys.}\ }\textbf {\bibinfo {volume} {86}},\ \bibinfo {pages} {995}
  (\bibinfo {year} {2014})}\BibitemShut {NoStop}%
\bibitem [{\citenamefont {Binder}\ \emph {et~al.}(2020)\citenamefont {Binder},
  \citenamefont {Egorov}, \citenamefont {Milchev},\ and\ \citenamefont
  {Nikoubashman}}]{Binder2020}%
  \BibitemOpen
  \bibfield  {author} {\bibinfo {author} {\bibfnamefont {K.}~\bibnamefont
  {Binder}}, \bibinfo {author} {\bibfnamefont {S.~A.}\ \bibnamefont {Egorov}},
  \bibinfo {author} {\bibfnamefont {A.}~\bibnamefont {Milchev}},\ and\ \bibinfo
  {author} {\bibfnamefont {A.}~\bibnamefont {Nikoubashman}},\ }\bibfield
  {title} {\bibinfo {title} {Polymer science overview},\ }\href@noop {}
  {\bibfield  {journal} {\bibinfo  {journal} {J. Phys.: Materials}\ }\textbf
  {\bibinfo {volume} {3}},\ \bibinfo {pages} {032008} (\bibinfo {year}
  {2020})}\BibitemShut {NoStop}%
\bibitem [{\citenamefont {Zhang}\ \emph {et~al.}(2015)\citenamefont {Zhang},
  \citenamefont {Gomez},\ and\ \citenamefont {Milner}}]{Zhang2015}%
  \BibitemOpen
  \bibfield  {author} {\bibinfo {author} {\bibfnamefont {W.}~\bibnamefont
  {Zhang}}, \bibinfo {author} {\bibfnamefont {E.~D.}\ \bibnamefont {Gomez}},\
  and\ \bibinfo {author} {\bibfnamefont {S.~T.}\ \bibnamefont {Milner}},\
  }\href@noop {} {\bibfield  {journal} {\bibinfo  {journal} {Macromolecules}\
  }\textbf {\bibinfo {volume} {48}},\ \bibinfo {pages} {1454} (\bibinfo {year}
  {2015})}\BibitemShut {NoStop}%
\bibitem [{\citenamefont {Botiz}\ and\ \citenamefont
  {Darling}(2010)}]{Botiz2010}%
  \BibitemOpen
  \bibfield  {author} {\bibinfo {author} {\bibfnamefont {I.}~\bibnamefont
  {Botiz}}\ and\ \bibinfo {author} {\bibfnamefont {S.~B.}\ \bibnamefont
  {Darling}},\ }\href@noop {} {\bibfield  {journal} {\bibinfo  {journal}
  {Materials Today}\ }\textbf {\bibinfo {volume} {13}},\ \bibinfo {pages} {42}
  (\bibinfo {year} {2010})}\BibitemShut {NoStop}%
\bibitem [{\citenamefont {Smith}\ \emph {et~al.}(1992)\citenamefont {Smith},
  \citenamefont {Finzi},\ and\ \citenamefont {Bustamante}}]{Smith1992}%
  \BibitemOpen
  \bibfield  {author} {\bibinfo {author} {\bibfnamefont {S.~B.}\ \bibnamefont
  {Smith}}, \bibinfo {author} {\bibfnamefont {L.}~\bibnamefont {Finzi}},\ and\
  \bibinfo {author} {\bibfnamefont {C.}~\bibnamefont {Bustamante}},\
  }\href@noop {} {\bibfield  {journal} {\bibinfo  {journal} {Science}\ }\textbf
  {\bibinfo {volume} {258}},\ \bibinfo {pages} {1122} (\bibinfo {year}
  {1992})}\BibitemShut {NoStop}%
\bibitem [{\citenamefont {Smith}\ \emph {et~al.}(1996)\citenamefont {Smith},
  \citenamefont {Cui},\ and\ \citenamefont {Bustamante}}]{Smith1996}%
  \BibitemOpen
  \bibfield  {author} {\bibinfo {author} {\bibfnamefont {S.~B.}\ \bibnamefont
  {Smith}}, \bibinfo {author} {\bibfnamefont {Y.}~\bibnamefont {Cui}},\ and\
  \bibinfo {author} {\bibfnamefont {C.}~\bibnamefont {Bustamante}},\
  }\href@noop {} {\bibfield  {journal} {\bibinfo  {journal} {Science}\ }\textbf
  {\bibinfo {volume} {271}},\ \bibinfo {pages} {795} (\bibinfo {year}
  {1996})}\BibitemShut {NoStop}%
\bibitem [{\citenamefont {Wiggins}\ \emph {et~al.}(2006)\citenamefont
  {Wiggins}, \citenamefont {van~der Heijden}, \citenamefont {Moreno-Herrero},
  \citenamefont {Spakowitz}, \citenamefont {Phillips}, \citenamefont {Widom},
  \citenamefont {Dekker},\ and\ \citenamefont {Nelson}}]{Wiggins2006}%
  \BibitemOpen
  \bibfield  {author} {\bibinfo {author} {\bibfnamefont {P.~A.}\ \bibnamefont
  {Wiggins}}, \bibinfo {author} {\bibfnamefont {T.}~\bibnamefont {van~der
  Heijden}}, \bibinfo {author} {\bibfnamefont {F.}~\bibnamefont
  {Moreno-Herrero}}, \bibinfo {author} {\bibfnamefont {A.}~\bibnamefont
  {Spakowitz}}, \bibinfo {author} {\bibfnamefont {R.}~\bibnamefont {Phillips}},
  \bibinfo {author} {\bibfnamefont {J.}~\bibnamefont {Widom}}, \bibinfo
  {author} {\bibfnamefont {C.}~\bibnamefont {Dekker}},\ and\ \bibinfo {author}
  {\bibfnamefont {P.~C.}\ \bibnamefont {Nelson}},\ }\href@noop {} {\bibfield
  {journal} {\bibinfo  {journal} {Nature Nanotechnology}\ }\textbf {\bibinfo
  {volume} {1}},\ \bibinfo {pages} {137} (\bibinfo {year} {2006})}\BibitemShut
  {NoStop}%
\bibitem [{\citenamefont {Mazur}(2007)}]{Mazur2007}%
  \BibitemOpen
  \bibfield  {author} {\bibinfo {author} {\bibfnamefont {A.~K.}\ \bibnamefont
  {Mazur}},\ }\href@noop {} {\bibfield  {journal} {\bibinfo  {journal} {Phys.
  Rev. Lett.}\ }\textbf {\bibinfo {volume} {98}},\ \bibinfo {pages} {218102}
  (\bibinfo {year} {2007})}\BibitemShut {NoStop}%
\bibitem [{\citenamefont {Midya}\ \emph {et~al.}(2019)\citenamefont {Midya},
  \citenamefont {Egorov}, \citenamefont {Binder},\ and\ \citenamefont
  {Nikoubashman}}]{Midya2019}%
  \BibitemOpen
  \bibfield  {author} {\bibinfo {author} {\bibfnamefont {J.}~\bibnamefont
  {Midya}}, \bibinfo {author} {\bibfnamefont {S.~A.}\ \bibnamefont {Egorov}},
  \bibinfo {author} {\bibfnamefont {K.}~\bibnamefont {Binder}},\ and\ \bibinfo
  {author} {\bibfnamefont {A.}~\bibnamefont {Nikoubashman}},\ }\bibfield
  {title} {\bibinfo {title} {Phase behavior of flexible and semiflexible
  polymers in solvents of varying quality},\ }\href@noop {} {\bibfield
  {journal} {\bibinfo  {journal} {J. Chem. Phy.}\ }\textbf {\bibinfo {volume}
  {151}},\ \bibinfo {pages} {034902} (\bibinfo {year} {2019})}\BibitemShut
  {NoStop}%
\bibitem [{\citenamefont {Kuhn}(1934)}]{Kuhn1934}%
  \BibitemOpen
  \bibfield  {author} {\bibinfo {author} {\bibfnamefont {W.}~\bibnamefont
  {Kuhn}},\ }\bibfield  {title} {\bibinfo {title} {Über die gestalt
  fadenförmiger moleküle in lösungen},\ }\href@noop {} {\bibfield  {journal}
  {\bibinfo  {journal} {Kolloid-Zeitschrift}\ }\textbf {\bibinfo {volume}
  {68}},\ \bibinfo {pages} {2} (\bibinfo {year} {1934})}\BibitemShut {NoStop}%
\bibitem [{\citenamefont {Kratky}\ and\ \citenamefont
  {Porod}(1949)}]{Kratky1949}%
  \BibitemOpen
  \bibfield  {author} {\bibinfo {author} {\bibfnamefont {O.}~\bibnamefont
  {Kratky}}\ and\ \bibinfo {author} {\bibfnamefont {G.}~\bibnamefont {Porod}},\
  }\href@noop {} {\bibfield  {journal} {\bibinfo  {journal} {Recueil des
  Travaux Chimiques des Pays-Bas}\ }\textbf {\bibinfo {volume} {68}},\ \bibinfo
  {pages} {1106} (\bibinfo {year} {1949})}\BibitemShut {NoStop}%
\bibitem [{\citenamefont {Liao}\ \emph {et~al.}(2020)\citenamefont {Liao},
  \citenamefont {Purohit},\ and\ \citenamefont {Gopinath}}]{Liao2020}%
  \BibitemOpen
  \bibfield  {author} {\bibinfo {author} {\bibfnamefont {X.}~\bibnamefont
  {Liao}}, \bibinfo {author} {\bibfnamefont {P.~K.}\ \bibnamefont {Purohit}},\
  and\ \bibinfo {author} {\bibfnamefont {A.}~\bibnamefont {Gopinath}},\
  }\bibfield  {title} {\bibinfo {title} {Extensions of the worm-like-chain
  model to tethered active filaments under tension},\ }\href@noop {} {\bibfield
   {journal} {\bibinfo  {journal} {J. Chem. Phys.}\ }\textbf {\bibinfo {volume}
  {153}},\ \bibinfo {pages} {194901} (\bibinfo {year} {2020})}\BibitemShut
  {NoStop}%
\bibitem [{\citenamefont {Milstein}\ and\ \citenamefont
  {Meiners}(2013)}]{Milstein2013}%
  \BibitemOpen
  \bibfield  {author} {\bibinfo {author} {\bibfnamefont {J.~N.}\ \bibnamefont
  {Milstein}}\ and\ \bibinfo {author} {\bibfnamefont {J.-C.}\ \bibnamefont
  {Meiners}},\ }\bibfield  {title} {\bibinfo {title} {Worm-like chain (wlc)
  model},\ }in\ \href@noop {} {\emph {\bibinfo {booktitle} {Encyclopedia of
  Biophysics}}}\ (\bibinfo  {publisher} {Springer},\ \bibinfo {address}
  {Berlin, Heidelberg},\ \bibinfo {year} {2013})\ pp.\ \bibinfo {pages}
  {2757--2760}\BibitemShut {NoStop}%
\bibitem [{\citenamefont {Cinacchi}\ and\ \citenamefont
  {De~Gaetani}(2008)}]{Cinacchi2008}%
  \BibitemOpen
  \bibfield  {author} {\bibinfo {author} {\bibfnamefont {G.}~\bibnamefont
  {Cinacchi}}\ and\ \bibinfo {author} {\bibfnamefont {L.}~\bibnamefont
  {De~Gaetani}},\ }\bibfield  {title} {\bibinfo {title} {Phase behavior of
  wormlike rods},\ }\href@noop {} {\bibfield  {journal} {\bibinfo  {journal}
  {Phys. Rev. E}\ }\textbf {\bibinfo {volume} {77}},\ \bibinfo {pages} {051705}
  (\bibinfo {year} {2008})}\BibitemShut {NoStop}%
\bibitem [{\citenamefont {Heussinger}\ \emph {et~al.}(2007)\citenamefont
  {Heussinger}, \citenamefont {Bathe},\ and\ \citenamefont
  {Frey}}]{Heussinger2007}%
  \BibitemOpen
  \bibfield  {author} {\bibinfo {author} {\bibfnamefont {C.}~\bibnamefont
  {Heussinger}}, \bibinfo {author} {\bibfnamefont {M.}~\bibnamefont {Bathe}},\
  and\ \bibinfo {author} {\bibfnamefont {E.}~\bibnamefont {Frey}},\ }\bibfield
  {title} {\bibinfo {title} {Statistical mechanics of semiflexible bundles of
  wormlike polymer chains},\ }\href@noop {} {\bibfield  {journal} {\bibinfo
  {journal} {Phys. Rev. Lett.}\ }\textbf {\bibinfo {volume} {99}},\ \bibinfo
  {pages} {048101} (\bibinfo {year} {2007})}\BibitemShut {NoStop}%
\bibitem [{\citenamefont {Samuel}\ and\ \citenamefont
  {Sinha}(2002)}]{Samuel2002}%
  \BibitemOpen
  \bibfield  {author} {\bibinfo {author} {\bibfnamefont {J.}~\bibnamefont
  {Samuel}}\ and\ \bibinfo {author} {\bibfnamefont {S.}~\bibnamefont {Sinha}},\
  }\bibfield  {title} {\bibinfo {title} {Elasticity of semiflexible polymers},\
  }\href@noop {} {\bibfield  {journal} {\bibinfo  {journal} {Phys. Rev. E}\
  }\textbf {\bibinfo {volume} {66}},\ \bibinfo {pages} {050801} (\bibinfo
  {year} {2002})}\BibitemShut {NoStop}%
\bibitem [{\citenamefont {Marantan}\ and\ \citenamefont
  {Mahadevan}(2018)}]{Marantan2018}%
  \BibitemOpen
  \bibfield  {author} {\bibinfo {author} {\bibfnamefont {A.}~\bibnamefont
  {Marantan}}\ and\ \bibinfo {author} {\bibfnamefont {L.}~\bibnamefont
  {Mahadevan}},\ }\href@noop {} {\bibfield  {journal} {\bibinfo  {journal} {Am.
  J. Phys.}\ }\textbf {\bibinfo {volume} {86}},\ \bibinfo {pages} {86}
  (\bibinfo {year} {2018})}\BibitemShut {NoStop}%
\bibitem [{\citenamefont {Fiasconaro}\ and\ \citenamefont
  {Falo}(2023)}]{Fiasconaro2023}%
  \BibitemOpen
  \bibfield  {author} {\bibinfo {author} {\bibfnamefont {A.}~\bibnamefont
  {Fiasconaro}}\ and\ \bibinfo {author} {\bibfnamefont {F.}~\bibnamefont
  {Falo}},\ }\bibfield  {title} {\bibinfo {title} {Elastic traits of the
  extensible discrete wormlike chain model},\ }\href@noop {} {\bibfield
  {journal} {\bibinfo  {journal} {Phys. Rev. E}\ }\textbf {\bibinfo {volume}
  {107}},\ \bibinfo {pages} {024501} (\bibinfo {year} {2023})}\BibitemShut
  {NoStop}%
\bibitem [{\citenamefont {Ghosh}\ \emph {et~al.}(2009)\citenamefont {Ghosh},
  \citenamefont {Singh},\ and\ \citenamefont {Sain}}]{Ghosh2009}%
  \BibitemOpen
  \bibfield  {author} {\bibinfo {author} {\bibfnamefont {S.~K.}\ \bibnamefont
  {Ghosh}}, \bibinfo {author} {\bibfnamefont {K.}~\bibnamefont {Singh}},\ and\
  \bibinfo {author} {\bibfnamefont {A.}~\bibnamefont {Sain}},\ }\bibfield
  {title} {\bibinfo {title} {Effect of intrinsic curvature on semiflexible
  polymers},\ }\href@noop {} {\bibfield  {journal} {\bibinfo  {journal} {Phys.
  Rev. E}\ }\textbf {\bibinfo {volume} {80}},\ \bibinfo {pages} {051904}
  (\bibinfo {year} {2009})}\BibitemShut {NoStop}%
\bibitem [{\citenamefont {Wiggins}\ \emph {et~al.}(2005)\citenamefont
  {Wiggins}, \citenamefont {Phillips},\ and\ \citenamefont
  {Nelson}}]{Wiggins2005}%
  \BibitemOpen
  \bibfield  {author} {\bibinfo {author} {\bibfnamefont {P.~A.}\ \bibnamefont
  {Wiggins}}, \bibinfo {author} {\bibfnamefont {R.}~\bibnamefont {Phillips}},\
  and\ \bibinfo {author} {\bibfnamefont {P.~C.}\ \bibnamefont {Nelson}},\
  }\bibfield  {title} {\bibinfo {title} {Exact theory of kinkable elastic
  polymers},\ }\href@noop {} {\bibfield  {journal} {\bibinfo  {journal} {Phys.
  Rev. E}\ }\textbf {\bibinfo {volume} {71}},\ \bibinfo {pages} {021909}
  (\bibinfo {year} {2005})}\BibitemShut {NoStop}%
\bibitem [{\citenamefont {Lamura}\ \emph {et~al.}(2001)\citenamefont {Lamura},
  \citenamefont {Burkhardt},\ and\ \citenamefont {Gompper}}]{Lamura2001}%
  \BibitemOpen
  \bibfield  {author} {\bibinfo {author} {\bibfnamefont {A.}~\bibnamefont
  {Lamura}}, \bibinfo {author} {\bibfnamefont {T.~W.}\ \bibnamefont
  {Burkhardt}},\ and\ \bibinfo {author} {\bibfnamefont {G.}~\bibnamefont
  {Gompper}},\ }\href@noop {} {\bibfield  {journal} {\bibinfo  {journal} {Phys.
  Rev. E}\ }\textbf {\bibinfo {volume} {64}},\ \bibinfo {pages} {061801}
  (\bibinfo {year} {2001})}\BibitemShut {NoStop}%
\bibitem [{\citenamefont {Nikoubashman}(2021)}]{Nikoubashman2021}%
  \BibitemOpen
  \bibfield  {author} {\bibinfo {author} {\bibfnamefont {A.}~\bibnamefont
  {Nikoubashman}},\ }\bibfield  {title} {\bibinfo {title} {Ordering, phase
  behavior, and correlations of semiflexible polymers in confinement},\
  }\href@noop {} {\bibfield  {journal} {\bibinfo  {journal} {J. Chem. Phys.}\
  }\textbf {\bibinfo {volume} {154}},\ \bibinfo {pages} {090901} (\bibinfo
  {year} {2021})}\BibitemShut {NoStop}%
\bibitem [{\citenamefont {Bakhti}(2026)}]{Bakhti2026}%
  \BibitemOpen
  \bibfield  {author} {\bibinfo {author} {\bibfnamefont {B.}~\bibnamefont
  {Bakhti}},\ }\bibfield  {title} {\bibinfo {title} {Resolving the short-{DNA}
  looping anomaly with anharmonic bending in the discrete wormlike chain}}
  (\bibinfo {year} {2026}),\ \bibinfo {note} {to be published}\BibitemShut
  {NoStop}%
\end{thebibliography}%

\end{document}